\documentclass[authoryear,preprint,12pt]{elsarticle}

\usepackage{graphicx}

\usepackage{amssymb}

\usepackage{setspace}
\usepackage{threeparttable}

\journal{Advances in Space Research}

\begin{document}

\begin{frontmatter}

\title{Solar Magnetic Feature Detection and Tracking for Space Weather Monitoring}

\author{P.A. Higgins\corref{cor1}}
\cortext[cor1]{Corresponding Authors}
\ead{pohuigin@gmail.com}

\author{P.T. Gallagher\corref{cor1}}
\author{R.T.J. McAteer}
\author{D.S. Bloomfield}

\address{Astrophysics Research Group, School of Physics, Trinity College Dublin, Dublin 2, Ireland}

\begin{abstract}

We present an automated system for detecting, tracking, and cataloging emerging active regions throughout their evolution and decay using \emph{SOHO} Michelson Doppler Interferometer (MDI) magnetograms. The \emph{SolarMonitor Active Region Tracking} (SMART) algorithm relies on consecutive image differencing to remove both quiet-Sun and transient magnetic features, and region-growing techniques to group flux concentrations into classifiable features. We determine magnetic properties such as region size, total flux, flux imbalance, flux emergence rate, Schrijver's $R$-value, $R^*$ (a modified version of $R$), and Falconer's measurement of non-potentiality. A persistence algorithm is used to associate developed active regions with emerging flux regions in previous measurements, and to track regions beyond the limb through multiple solar rotations. We find that the total number and area of magnetic regions on disk vary with the sunspot cycle. While sunspot numbers are a proxy to the solar magnetic field, SMART offers a direct diagnostic of the surface magnetic field and its variation over timescale of hours to years. SMART will form the basis of the active region extraction and tracking algorithm for the Heliophysics Integrated Observatory (HELIO). 


{\it }

\end{abstract}


\end{frontmatter}

\section{Introduction}

The automatic identification and characterization of solar features is of great importance to both solar activity monitoring and space weather operations. This has become a particular issue due to the high spatial and temporal resolution solar imagers, such as those flown on the \emph{Project for On-Board Autonomy 2} (\emph{PROBA2}) and \emph{Solar Dynamics Observatory} (\emph{SDO}), which will force data providers to distribute subsets of their science products instead of the full image data set. Traditionally, solar feature catalogs were created by hand, using visual recognition to record the position, size, and other properties of features \citep[e.g.,][]{Car1854}. An early attempt to overcome this was SolarMonitor\footnote{See: http://www.SolarMonitor.org}\citep{Gal02}, which labels active regions (ARs) in solar images using National Oceanic and Atmospheric Administration (NOAA) numbers and locations cataloged by the NOAA Space Weather Prediction Center.  More recently, researchers have begun to catalog features using automated methods. The European Grid of Solar Observations\footnote{See: http://www.egso.org}  \citep[EGSO;][]{Bently02}, for example, catalogs solar features using H$\alpha$ and Ca II K images and a neural network algorithm \citep{Zaretal05,ZarSch05}. 

One of the first applications of automated image processing techniques to AR identification is the Automated Region Selection Extraction algorithm \citep{McAt05a}. This algorithm creates a binary mask of features using a static noise threshold applied to a line-of-sight (LOS) magnetogram. A sub-image is extracted, centered on the pixel with the highest value. Closed contours enclosing an area centered on the seed are grouped as a region. The detected region is saved and removed from the magnetogram. The pixel of the next highest value is selected and the process repeated. Some saved regions are associated with NOAA cataloged regions which may be tracked across the disk. More recently,  \citet{LaBon07} extract ARs using full-disk magnetograms that are smoothed by roughly one supergranule diameter. Region candidates are tested for bipolar flux and east-west orientation. A dynamic noise threshold is calculated using the median of average magnetic field values for a series of annuli centered on the AR candidate. The AR boundary is chosen by comparing the average magnetic field values of smaller annuli with the calculated noise threshold. Using annuli to test for region boundaries allows one to isolate ARs from large AR complexes since the dynamic noise threshold will be set relative to the surrounding regions.

An alternative to solely identifying ARs using their magnetic signatures was discussed in a series of papers by \citet{QaCo05} and \citet{CoQa08, CoQa09}. In their hybrid extraction algorithm, sunspots detected in white-light images are grouped using feature boundaries extracted from magnetograms. Both forms of data are segmented using dynamic thresholding. White-light candidates coinciding with magnetogram candidates are grouped using growing circles, while a neural network is used to determine which candidates to retain and how to group them. This system has the advantage that it compares well the NOAA AR identification scheme, but it does not give any insight into AR properties thought to be related to flaring (e.g., horizontal B-field gradients, total flux, fractal dimension, etc.)

A number of algorithms have been developed to measure AR magnetic characteristics postulated to be related to flaring: \citet{Gal02} measure gradients in the magnetic field of ARs; \citet{McAt05b} establish an AR fractal dimension lower limit of 1.2 for M- and X-class flares to occur; \citet{GeRu07} calculate the magnetic connectivity between fragments of an AR;  \citet{Con08,Con10} measure the multifractal nature of AR flux; \citet{Hew08} determine the multiscale power-law index;  \citet{Fal08} establish a gauge of AR non-potentiality; and \citet{JieZ09} determine basic field properties and the degree of AR polarity (bipole, quadrupole, etc.). The overall aim of these algorithms is to extract a physically-motivated measure of the characteristics of a region, subsequently using this information to better understand the fundamental physics of ARs, and to relate the properties of AR magnetic fields to their flaring potential. This is essential to building an automated AR monitoring and flare forecasting system as discussed in \citet{McAt09}. 

In this paper we present a new algorithm, the SolarMonitor Active Region Tracking (SMART) algorithm, which will form the basis of AR identification for the Heliophysics Integrated Observatory\footnote{See: http://www.helio-vo.org} (HELIO).  SMART combines extraction techniques with AR magnetic property determinations (Section~\ref{algorithm}), region tracking, and cataloging (Section~\ref{tracking}). A cross comparison of SMART and NOAA detections as well as a discussion of errors in property measurements and feature tracking test cases are presented in Section~\ref{results}. Our conclusions and prospects for future work are then given in Section~\ref{disc_conc}.

\section{Feature Extraction}\label{algorithm}

\begin{figure}
\begin{center}
\includegraphics*[width=8cm,angle=0]{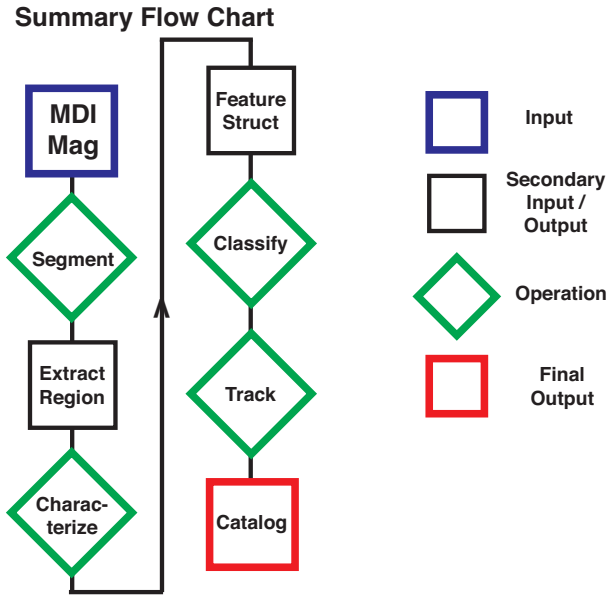}
\end{center}
\caption{Flow chart summarizing the SMART algorithm processing method.}\label{flowsummary}
\end{figure}

\indent The SMART method of operation is summarized in Figure~\ref{flowsummary}. Initially, magnetograms are segmented into individual feature masks (Section~\ref{dataproc}). A characterization algorithm is then run on each extracted region to determine feature properties (Section~\ref{magprop}). These region properties are subsequently used to classify the form of solar features (Section~\ref{classify}).
The final output is a set of data structures for each magnetogram, including each feature present. The following sub-sections provide details on the operations outlined above.

\subsection{Segmentation}\label{dataproc}

\begin{figure}[!ht]
\begin{center}
\includegraphics*[width=10cm,angle=0]{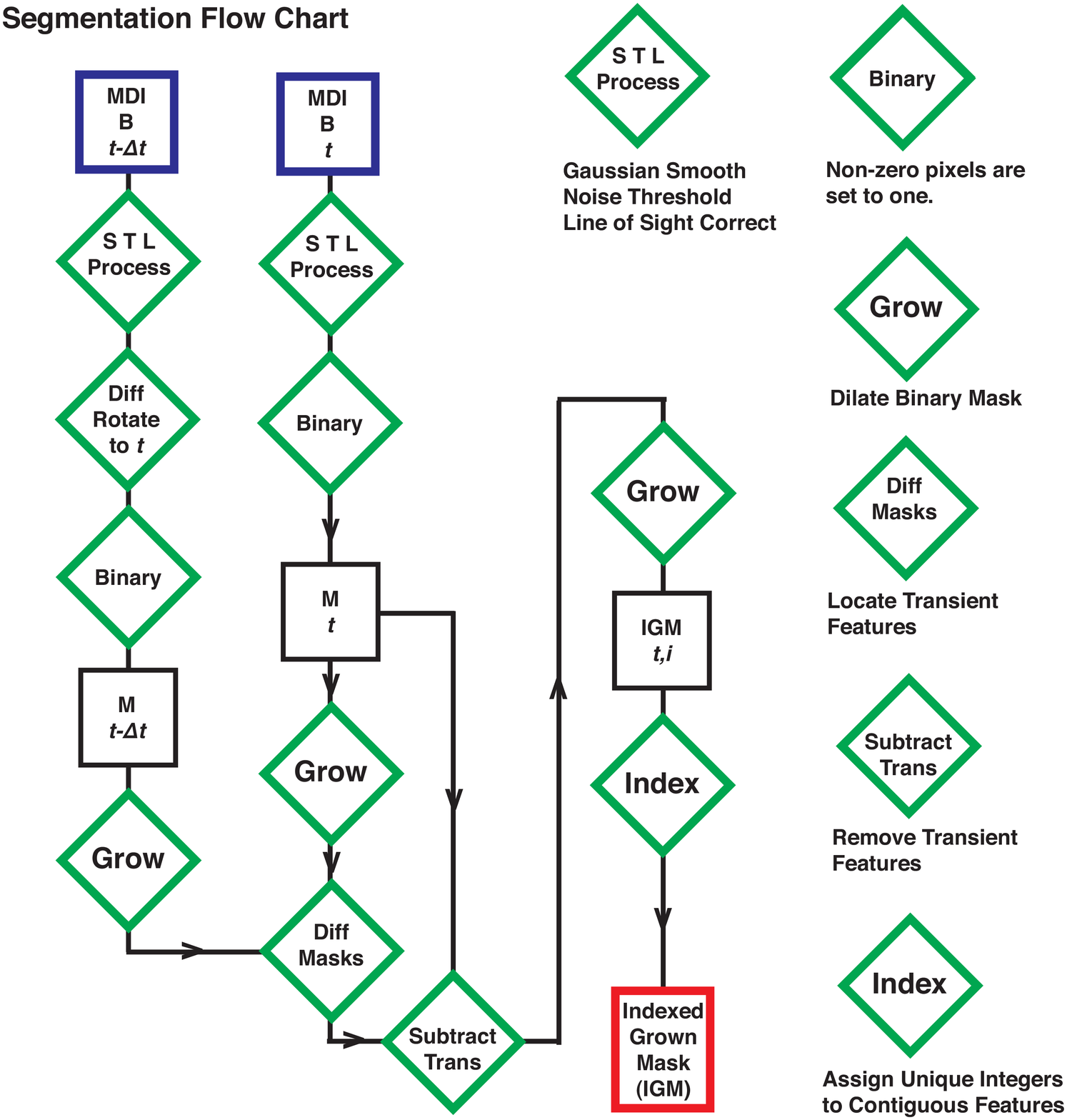}
\end{center}
\caption{Flow chart summarizing the magnetogram segmentation method.}\label{segment}
\end{figure}

\begin{figure}[!ht]
\begin{center}
\includegraphics*[width=8cm,angle=0]{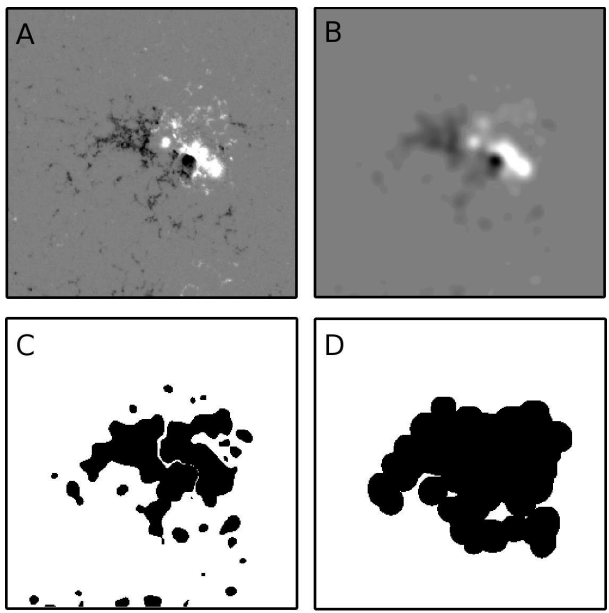}
\end{center}
\caption{Processing steps for an example feature extraction on 25 November 2003. A) Calibrated megnetogram $B_{t}$ clipped to $\pm1000$~G and cropped around NOAA 10507. B) $B_{t}$ with gaussian smoothing and noise thresholding. C) Mask ($M_{f,t}$) with transient filtering and area threshold of $50$~pixels. D) Final indexed grown feature mask, $IGM_{t,i}$.}\label{processing}
\end{figure}

\indent The segmentation process depicted in Figure~\ref{segment} begins with two consecutive \emph{Solar and Heliospheric Observatory} (\emph{SOHO})/Michelson Doppler Interferometer
\citep[MDI;][]{Sher95} full-disk, line-of-sight (LOS), level 1.8 magnetograms. 
Nominally these are $96$~minutes apart, but there are sporadic gaps in the MDI data set (only rarely is there an entire day with no data). We use two magnetograms recorded close in time to remove transient features and extract time-dependent properties. The magnetogram of interest (Figure~\ref{processing} A) is denoted as $B_{t}$ and the previous magnetogram as $B_{t-\Delta t}$. If $\Delta t$ is greater than one day, the detections are discarded.

\indent Magnetograms are first checked for problems using properties extracted from the Flexible Image Transport System (FITS) data file headers, such as the spacecraft roll angle and the number of missing pixel values. Magnetograms are rotated as necessary, so that solar north points up, using nearest neighbor sampling interpolation, while those with missing values are discarded. A solar energetic particle (SEP) event which occurs during a magnetogram exposure results in many bright pixels scattered about the image. This does not interfere with the AR detection, as the bright pixels are smoothed out, but can affect magnetic property determinations. 

\indent We first apply smoothing, a noise threshold, and a LOS correction, respectively, to the data (Figure~\ref{processing} B). This set of operations is represented by \emph{STL Process} in Figure \ref{segment}. The smoothing operation is necessary to remove ephemeral regions that have size scales on the order of $10$~Mm \citep{HH01}, which corresponds to $7$~MDI~pixels at disk center.
To this end, $B_{t-\Delta t}$ and $B_{t}$ are convolved with a $10\times 10$~pixel$^2$ kernel containing a 2D gaussian with a full-width at half-maximum (FWHM) of $5$~pixels. 

\begin{figure}[!ht]
\label{cycledist}
\begin{center}
\includegraphics*[width=12cm,angle=0]{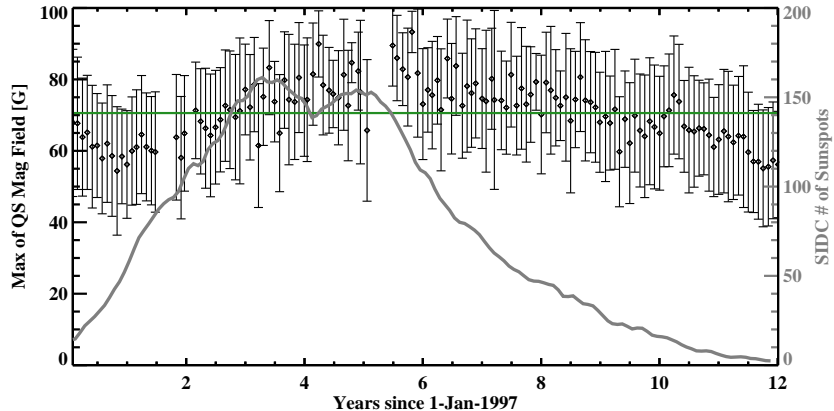}
\end{center}
\caption{The maximum of quiet-Sun magnetic field values over solar cycle 23. Each point is the mean maximum value for a month of magnetograms (nominally two per day, but less for particularly active periods). The error bars are the standard deviations of each month's set of values. The continuous gray line is the smoothed, monthly sunspot number from Solar Influences Data Analysis Center (SIDC; http://sidc.oma.be).}\label{cycle_sig}
\end{figure}

\begin{figure}[!ht]
\label{betadist}
\begin{center}
\includegraphics*[width=10cm,angle=0]{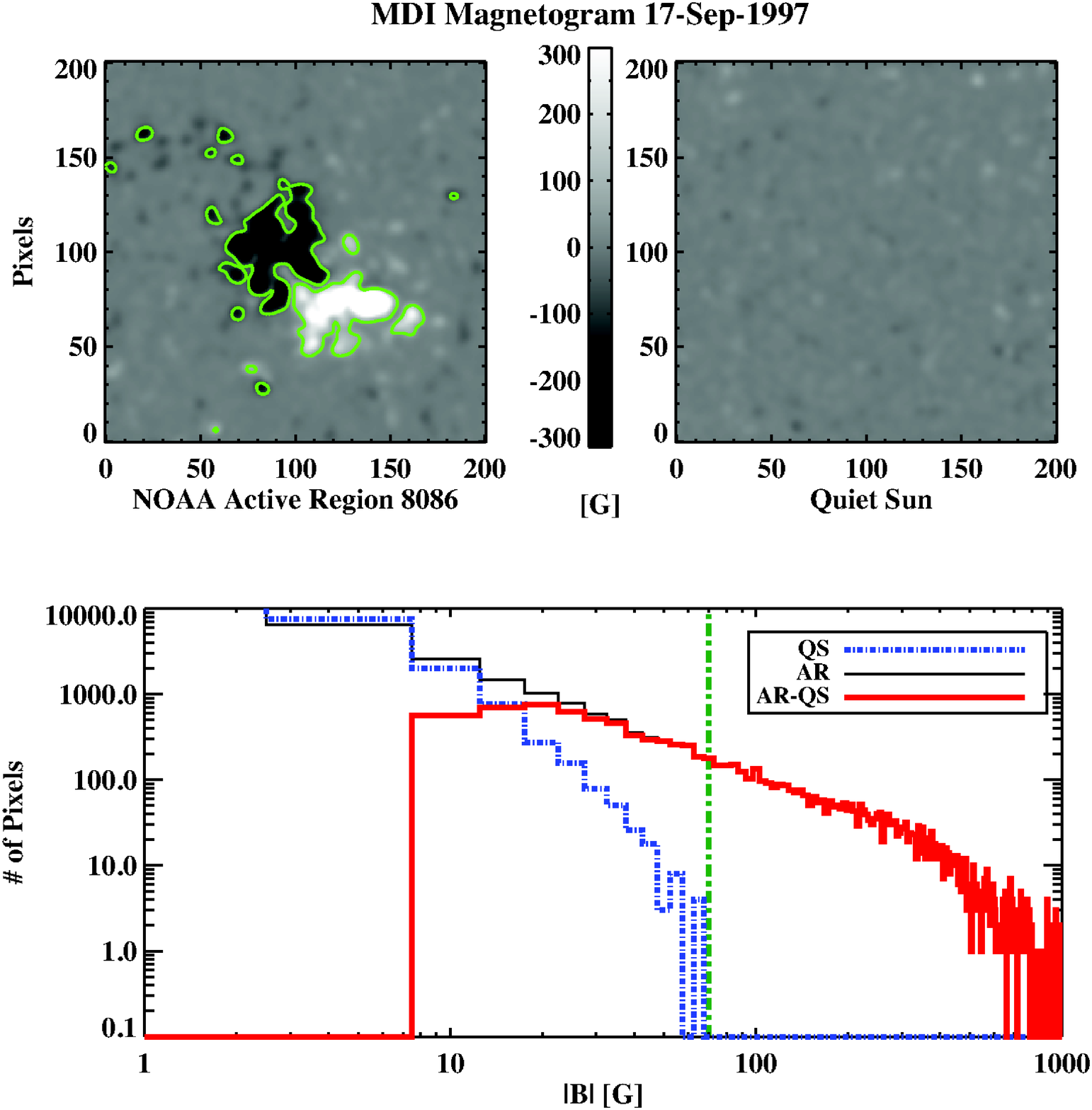}
\end{center}
\caption{A comparison of magnetic field value distributions for a quiet Sun and solar feature region. \emph{Top left}: Magnetogram of NOAA 8086, gaussian smoothed using a FWHM of 5 pixels. \emph{Top right}: A nearby region of quiet Sun in the same full-disk image. \emph{Bottom}: The feature and quiet-Sun unsigned magnetic field distributions. The thick red line is the difference between the quiet-Sun and AR distributions and the vertical dash-dotted line denotes $70$~G.}\label{ar_qs_dist}
\end{figure}

\indent We use a static threshold to remove the background. Figure~\ref{cycle_sig} shows the variation in the monthly averages of maximum values of quiet-Sun (QS) magnetic field recorded throughout cycle 23. The maximum value varies by roughly $5$~G over the cycle which is less than the monthly standard deviation of these maxima, so a static threshold is acceptable. The mean of the maximum unsigned QS magnetic field values is $\sim$$70$~G. Figure~\ref{ar_qs_dist} shows a smoothed AR and nearby QS region contoured at $\pm 70$~G. The histogram shows the distributions of magnetic field values for the AR and QS regions, including the difference between the two distributions. Thresholding at the  $\pm 70$~G level removes small features which have been smoothed out by the gaussian convolution but maintains extended strong-field features, such as bipolar and plage regions. Pixels in $B_{t-\Delta t}$ and $B_{t}$ with absolute values less than $70$~G are zeroed.

\indent In the case where magnetic fields are primarily vertical to the solar surface, the LOS component of the field is reduced toward the limb. As such, a feature with the same magnetic field strength and orientation with respect to the solar surface will appear lower in magnitude when located toward the solar limb than at disk center. This LOS effect is corrected at each MDI pixel using a cosine correction factor \citep{McAt05a}.
After this stage, $B_{t-\Delta t}$ data is differentially rotated to time $t$ to correct for feature motions due to solar rotation using the latitudinal dependence derived in \citet{HowHarFor90}.

\begin{table}[!ht]
\begin{center}
\begin{tabular}{lcl}
\hline
Data Type&Identifier&Explanation\\
\hline
Feature Array&$B_{t,i}$ & extracted feature magnetogram \\
&$IGM_{t,i}$ & extracted feature mask \\ 
&$HG_{t}$ & heliographic position map \\
&$A_{cos,t,i}$ & $ \frac{IGM_{t,i}}{ \cos(HG_{t}} \times (\mbox{1.4 Mm}^{2}/\mbox{pixel})^{-1}$ \\
&${\Phi}_{t,i}$ & $B_{t}\times A_{cos,t,i}$ \\
&${\frac{d\Phi}{dt}}|_{t,i}$ & $\frac{(\mid B_{t}\mid - \mid B_{t-\Delta t}\mid )\times A_{cos,t,i}}{\Delta t}$\\
\hline
Property Value&$HG_{pos,t,i}$ & $\frac{\sum_{pix} ({B_{t,i}\times HG_{t}})}{\sum_{pix}{(HG_{t} \times IGM_{t,i})}}$ \\
&$B_{max,t,i}$ & maximum value of $B_{t,i}$ \\
&$B_{min,t,i}$ & minimum value of $B_{t,i}$ \\
&$B_{tot,t,i}$ & $\sum_{pix} {B_{t,i}}$ \\
&$B_{tot\mbox{ }uns,t,i}$ & $\sum_{pix} {\mid B_{t,i}\mid}$ \\
&$\mu,\sigma^2,\gamma,\kappa$ & mean, variance, skewness, kurtosis \\
&$A_{tot,t,i}$ & $\sum_{pix} A_{cos,t,i}$ \\
&${\Phi}_{+,t,i}$ & $\sum_{pix} {({\Phi}_{t,i}>0)}$ \\
&${\Phi}_{-,t,i}$ & $\sum_{pix} {({\Phi}_{t,i}<0)}$ \\
&${\Phi}_{uns,t,i}$ & $\sum_{pix} {\mid {\Phi}_{t,i}\mid}$ \\
&${\Phi}_{imb,t,i}$ & $\frac{\mid({\Phi}_{+,t,i}-\mid {\Phi}_{-,t,i}\mid )\mid }{ \Phi_{uns,t,i}}$ \\
&$\frac{d\Phi}{dt}|_{net,t,i}$ & $\sum_{pix} {{\frac{d\Phi}{dt}}|_{t,i}}$ \\
\hline
\end{tabular}
\caption{Feature magnetic properties derived from characterization processing.}
\label{tablemagprop}
\end{center}
\end{table}

\indent The corrected magnetograms are made binary by setting all pixels with magnetic field values above the $\pm 70$~G threshold equal to one, yielding masks $M_{t-\Delta t}$ and $M_{t}$. Features consisting of less than $50$~pixels and those which are not present in both masks are removed by the following operations (Figure~\ref{processing} C). Firstly, each mask is dilated by $10$~pixels to allow for region expansion. Secondly, the binary masks are subtracted such that non-zero pixels in the difference mask identify features only occurring in $M_{t-\Delta t}$ or $M_{t}$. These transient features are subsequently removed from the un-grown version of $M_{t}$, which is then dilated by $10$~pixels to form $M_{f,t}$ (Figure~\ref{processing} D). Individual contiguous features in $M_{f,t}$ are indexed by assigning ascending integer values (beginning with one) in order of decreasing feature size. The segmentation output is an indexed grown mask ($IGM_{t}$), as shown by the thick red box in Figure~\ref{segment}.

\subsection{Characterization}\label{magprop}

\begin{figure}[!ht]
\begin{center}
\includegraphics*[width=12cm,angle=0]{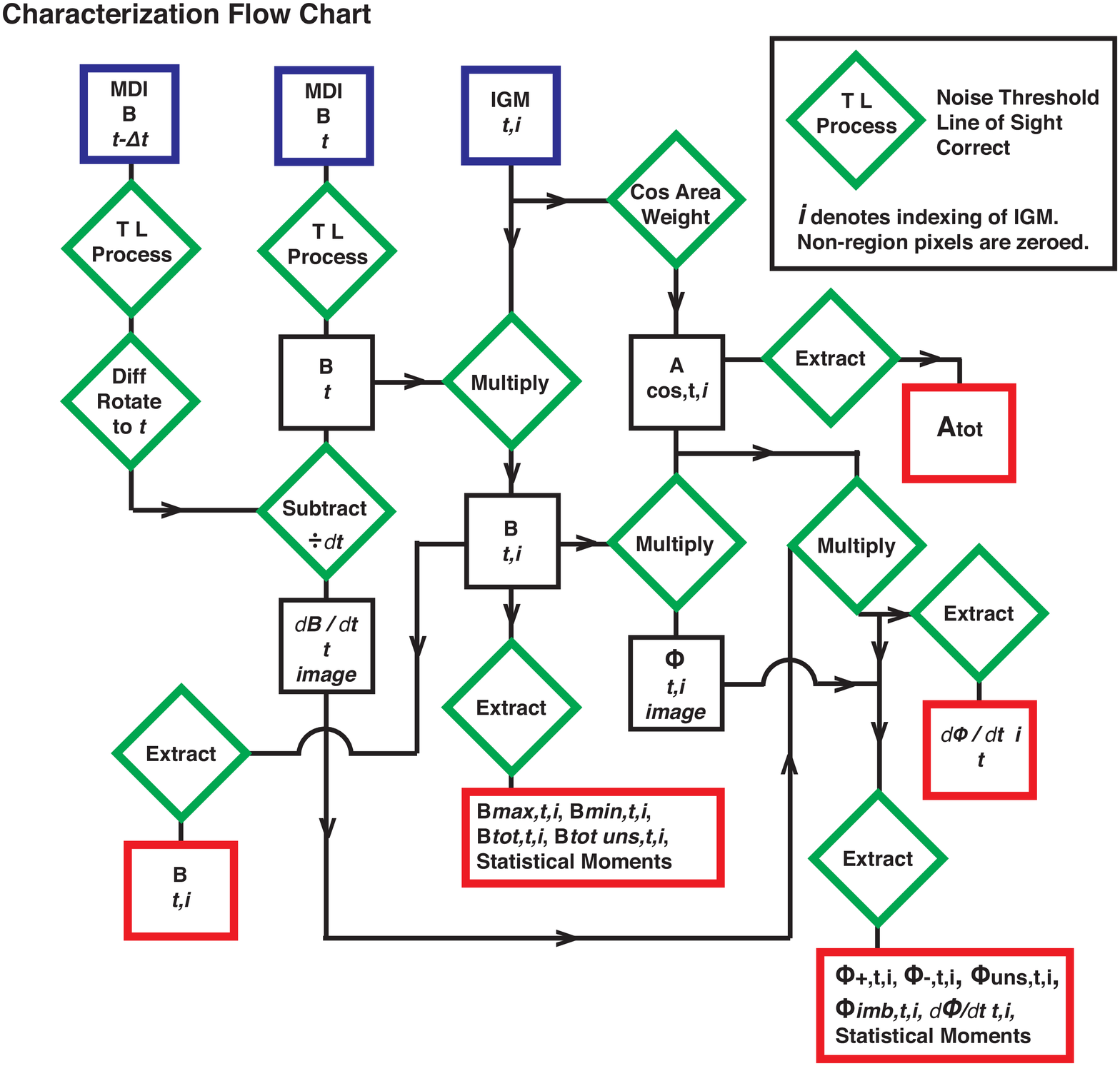}
\end{center}
\caption{Flow chart summarizing the feature magnetic property characterization method.}\label{flow_char}
\end{figure}

\indent The aim of SMART is to characterize ARs in a manner which does not make theoretical assumptions or require many observations of the same feature.  Our design is adaptable, so that the software may produce initial results in near-realtime for operational purposes, but allows the retrospective addition of complex property measurements (e.g., magnetic helicity). These requirements define criteria for the selection of initial property calculations. There are many AR properties that may be derived from magnetograms. A subset of these are derived from 96 minute LOS data and those output by SMART are included in Tables~1 and 2. 

\indent The SMART characterization process utilizes the feature mask retrieved by the methods outlined in the previous section, following the procedure detailed in Figure~\ref{flow_char}. The property measurements are derived from the magnetogram taken at time, $t$ which is processed in the manner detailed below. We subscript the mask containing all features by $i$ to extract a single feature mask, $IGM_{t,i}$. A cosine-weighted area map, $A_{cos,t,i}$  is derived which corrects pixel areas to solar surface area rather than plane-of-sky area, and is summed to yield total feature area, $A_{tot,t,i}$. 

\begin{table}[!ht]
\begin{center}
\begin{threeparttable}
\begin{tabular}{lcl}
\hline
Data Type&Identifier&Explanation\\
\hline
Feature Array&$M_{PSL,t,i}$ & polarity separation line mask \\
&$M_{PSL,thin,t,i}$ & thinned polarity separation line mask \\
\hline
Property Value&$L_{PSL,t,i}$ & $\sum_{pix}{M_{PSL,thin,t,i}}$ \\
&$L_{sg,t,i}$ & $L_{PSL,t,i} > 50$~G~Mm$^{-1}$ \\
&$R_{t,i}$ & $R$-value\tnote{1} \\ 
&$R^{*}_{t,i}$ & $\sum_{pix}{(M_{PSL,t,i} \ast Gauss_{2D})\times B_{t,i}} $ \\
&$WL_{sg,t,i}$ & non-potentiality gauge\tnote{2} \\ 
&$WL^{*}_{sg,t,i}$ & $\sum_{pix}{M_{PSL,t,i} \times \nabla B_{t,i}} $ \\
\hline
\end{tabular}
\begin{tablenotes} \footnotesize
\item[1] \citet{Shj07}
\item[2] \citet{Fal08}
\end{tablenotes}
\end{threeparttable}
\caption{Feature magnetic properties derived from polarity separation line characterization.}
\label{tablehighorder}
\end{center}
\end{table}

\indent Full-disk magnetograms $B_{t-\Delta t}$ and $B_{t}$ are processed as in Section \ref{dataproc} (thresholding, LOS correction, $B_{t-\Delta t}$ differentially rotated to time $t$), but without smoothing. Single features are extracted for magnetic property determination using the indexed grown mask,
\begin{equation}\label{eqn_ardata}{%
    B_{t,i} = {IGM_{t,i}} \times {B_{t}} \mbox{ \ ,}
}
\overfullrule 5pt
\end{equation}
yielding an array where all pixels but those in the feature are set to zero. 
The processed magnetograms  are subtracted and divided by their time separation to yield a map of the temporal change in field strength, ${{dB}/{dt}}|_{t}$, leading up to time $t$. This is combined with $A_{cos,t,i}$ to determine the flux emergence rate, ${{d\Phi}/{dt}}|_{t,i}$, of feature $i$.

\begin{figure}[!ht]
\begin{center}
\includegraphics*[width=10cm,angle=0]{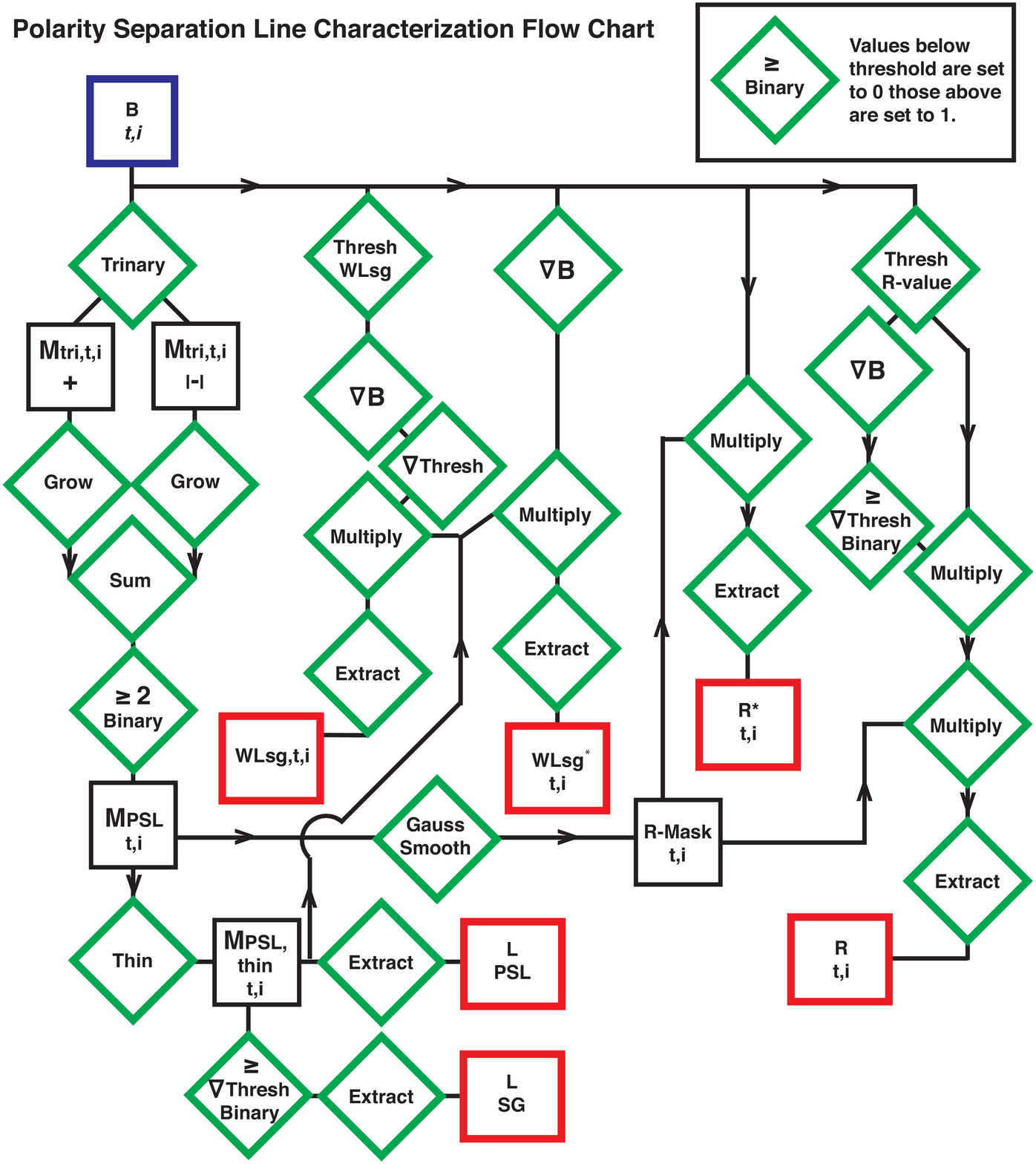}
\end{center}
\caption{Flow chart summarizing the quantities derived from feature polarity separation lines.}\label{flow_highorder}
\end{figure}

The extracted $B_{t,i}$ is used to extract other properties from feature $i$ (as detailed in Table~1) such as statistical moments of the magnetic field and the minimum and maximum magnetic field values ($B_{min,t,i}$ and $B_{max,t,i}$). $B_{t,i}$ is multiplied by $A_{cos,t,i}$ to derive the total positive, negative, and unsigned flux (${\Phi}_{+,t,i}$, ${\Phi}_{-,t,i}$, and ${\Phi}_{uns,t,i}$), the relative flux imbalance (${\Phi}_{imb,t,i}$), and the net flux emergence rate (${{d\Phi}/{dt}}|_{net,t,i}$).

\begin{figure}[!ht]
\begin{center}
\includegraphics*[width=8cm,angle=0]{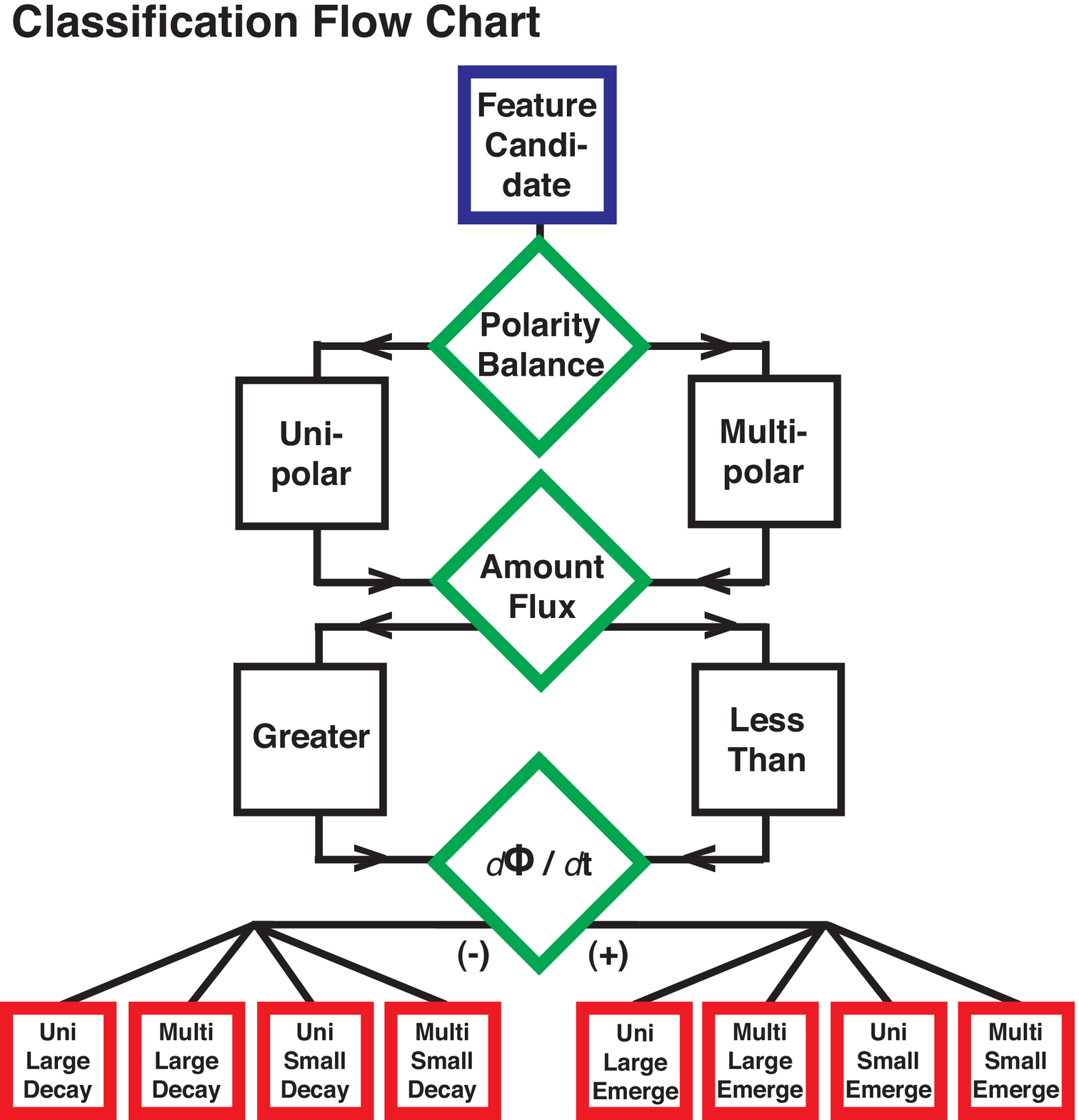}
\end{center}
\caption{Flow chart summarizing the feature classification method.}\label{flow_class}
\end{figure}

\indent The extracted feature magnetogram, $B_{t,i}$, is also used to derive properties based on the polarity separation line (PSL). Figure~\ref{flow_highorder} summarizes the extraction of feature properties related to PSLs and Table~2 lists the properties derived. Initially, the feature is segmented into its positive and negative components. These components are used to create a positive and negative mask, each of which is dilated by $4$~pixels. The two masks are summed and the region of mask overlap becomes the PSL binary mask, $M_{PSL,t,i}$. The algorithm then thins $M_{PSL,t,i}$ to one pixel ($M_{PSL,thin,t,i}$) and sums the non-zero pixels to determine the PSL length ($L_{PSL,t,i}$).  $L_{sg,t,i}$ is obtained by summing only those pixels which have $\nabla B_{t,i} > 50$~G~Mm$^{-1}$, where $\nabla B_{t,i}$ is calculated by numerical differentiation using 3-point Lagrangian interpolation. We also calculate the $R$-value ($R_{t,i}$) as presented in \citet{Shj07} and the $WL_{sg}$ gauge ($WL_{sg,t,i}$) as presented in \citet{Fal08}, both of which use specific gradient and magnetic field thresholding when extracting the PSL. 

\indent Finally, using $M_{PSL,t,i}$ we calculate $R^{*}_{t,i}$ which is a more sensitive version of the $R$-value, since it contains no gradient thresholding and the magnetic field threshold of $\pm70$~G is much lower than the $\pm150$~G used in \citet{Shj07}. The algorithm convolves $M_{PSL,t,i}$ with a $20\times20$~pixel$^2$ kernel containing a 2D gaussian with a FWHM of $10$~pixels, which is multiplied by $B_{t,i}$ and summed to achieve $R_{t,i}^{*}$. Similarly, an alternative of $WL_{sg,t,i}$, $WL^{*}_{sg,t,i}$ is calculated by applying the 
\citet{Fal08} method, but using a magnetic field 
threshold of $\pm70$~G and no gradient threshold.

\indent A set of data structures is created for each magnetogram including the above mentioned properties of each extracted feature (used for classification; Section~\ref{classify}) and the feature's heliographic location and time of measurement (used for feature tracking; Section~\ref{tracking}).

\subsection{Classification}\label{classify}

\indent At this stage the SMART algorithm has characterized the properties of each automatically extracted feature. The classification process uses these properties to discriminate between various feature types, which are saved in the algorithm output. Extracted features are initially grouped (as shown in Figure~\ref{flow_class}) into two catagories: 
features with a flux imbalance greater than 90\% are classified unipolar (U), while those having less than 90\% are classified multipolar (M). After polarity balance, the total unsigned magnetic flux ($\Phi_{uns,t,i}$) is tested. Features with $\Phi_{uns,t,i}$ greater than $10^{21}$~Mx are classified as large (L), while features with $\Phi_{uns,t,i}$ less than $10^{21}$~Mx are classified as small (S).
Finally, the sign of ${\frac{d\Phi}{dt}}|_{t,i}$ is tested to determine if features are increasing in flux (emerging, E) or decreasing in flux (decaying, D). The classification scheme results in eight possible feature classifications which are then also attributed to common magnetic feature designations: MLE and MLD are denoted evolving ARs; MSE and USE are denoted emerging flux concentration (EF); MSD, USD, and are denoted decaying flux concentrations (DF), and finally, ULE and ULD are denoted plage (PL). These common designations are also saved in the algorithm output, allowing one to make a quick assessment of which regions on disk are interesting from a monitoring point of view. For example, EFs may become ARs and evolving ARs may produce activity during their evolution, while PL and DF are not likely to produce activity.

\section{Tracking}\label{tracking}

Having detected various solar features, the SMART algorithm associates features across different time intervals. 
Spatial and temporal information is used to track features between consecutive images (Section~\ref{history}) and around the far-side of the disk between consecutive solar rotations (Sections~\ref{history2}). Features are then cataloged using the time of their first detection and their classification (Section~\ref{cataloging}). 

\subsection{Consecutive Images}\label{history}

\indent The set of features in a magnetogram is compared with the previous five magnetogram sets to associate previously catalogued features with the current set. Feature positions ($HG_{pos,t,i}$) are differentially rotated, using the latitudinal dependence derived in \citet{HowHarFor90}, to the same time $t$ and features matched when their heliographic separations are less than $5$~degrees. Features having one classification in previous sets may be associated with features having a different one in the current set. Thus, the SMART algorithm is capable of tracking possible ARs (MLE, MLD) back to their first emergence as an EF (MSE). Decaying features may also be associated with features previously denoted as possible ARs, allowing ARs to be followed through their final stages of evolution.
Fragmentation often occurs in these late stages which SMART allows for since it does not preclude multiple features from being associated with a single previous feature. If one feature splits into two, each resulting fragment will be associated with the original feature if the resulting fragment positions are within the matching threshold of the original. A letter is appended to the catalog name of each additional associated feature so that individual fragments may be differentiated. 

\subsection{Far-side Passage}\label{history2}

Features are tracked beyond the limb through multiple solar rotations to study their evolution from emergence to decay. We calculate the rotation period, $P_{rot,i}$, for each feature at time $t$, which depends on its heliographic latitude due to differential solar rotation. The feature position is compared to those in the five magnetogram sets centered on time $(t+t_{70})-P_{rot,i}$ using the method in the previous section. In this approach the feature position is essentially rotated to a longitude of $+70$~degrees then back one full solar rotation, where $t_{70}$ is the time taken for the feature to rotate to $70$~degrees heliographic longitude from its position at $t$. In this way the feature is constrained to have been previously detected just before west limb passage, which increases the efficiency of the algorithm.

\subsection{Cataloging}\label{cataloging}

There are two identifications recorded for each detected feature in a magnetogram at time, $t$. The first, $i$ is obtained from $IGM_{t}$ and denotes the two-digit size order of the feature. A feature within a single magnetogram is uniquely identified by $i$. The second identification is the static catalog name, YYYYMMDD.MG.NN, where YYYY is the four digit year, MM is the two digit month, and DD is the two digit day. The next two characters specify the feature type: MG denotes a photospheric magnetic feature. This scheme can be expanded to incorporate coronal holes (CH), filaments (FI), and transient features such as flares (FL) and coronal mass ejections (CE) in EUV images. Finally, NN is $i$ when the feature is given a static catalog name. This catalog name is determined once for each feature upon first detection, and is used for all measurements of the same feature as it is tracked through time. 


\begin{figure}[!ht]
\begin{center}
\includegraphics*[width=14cm,angle=0]{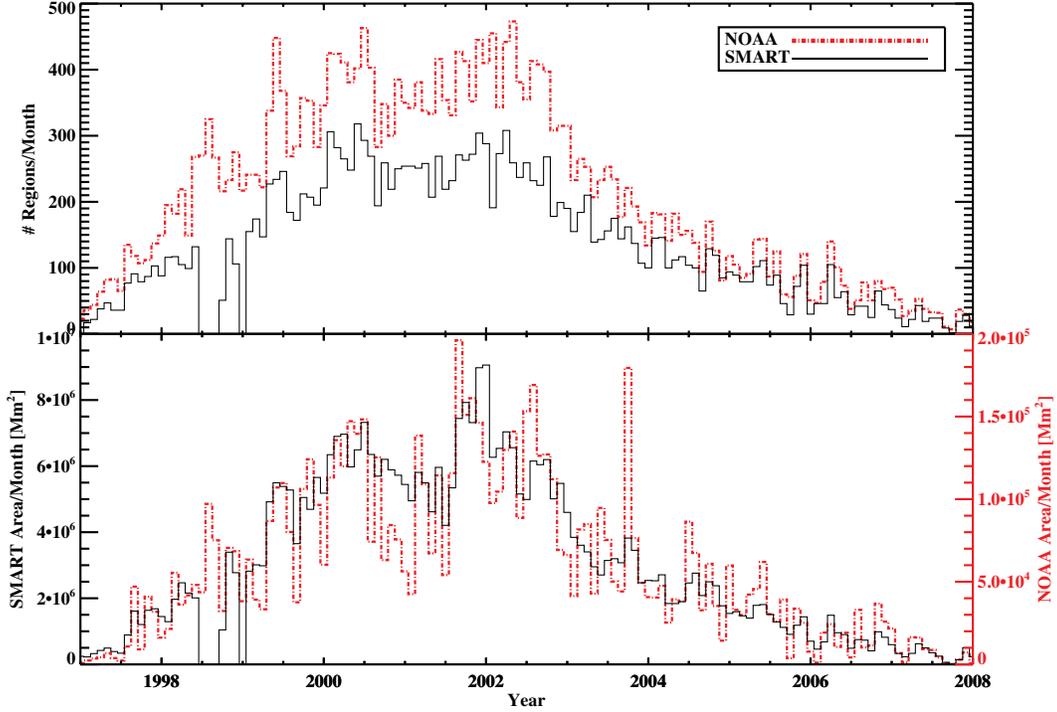}
\end{center}
\caption{A comparison of NOAA and SMART AR detections (binned by $1$~month) over cycle 23. The data gap in 1998 is due a the loss of communications with the SOHO spacecraft for several months.}\label{noaa_comp}
\end{figure}

\section{Results and Discussion}\label{results}

Figure~\ref{noaa_comp} summarizes a comparison of NOAA and SMART AR detections over the cycle 23, including numbers of detections and total feature area on disk. The top panel shows the total number of regions detected in each data set, arranged in monthly bins; the correlation coefficient between the (un-binned) daily data is $0.88$. We estimate the frequency of divergence between the detections using the ratio of NOAA to SMART AR daily detections: the ratio is between zero and one $6\%$, equal to one $22\%$, between one and two $60\%$, and greater than two $12\%$ of the time. We see a smaller number of SMART than NOAA AR detections $72\%$ of the time; the mean ratio of NOAA to SMART AR detections is $1.5$. This is likely due to the joining of two or more nearby sunspot groups by SMART, while NOAA identifies each individual sunspot group, regardless of proximity\footnote{NOAA may also detect very weak sunspots which may have a $\Phi_{uns,t,i}$ too small for designation as an AR by SMART.}. As such, SMART detections are representative of isolated magnetic systems, while NOAA detections represent a feature recognition approach. Additionally, NOAA records detections by eye, and only if they are visible in intensity data (i.e., if there is a magnetic flux concentration with no sunspot SMART may detect a region when NOAA does not). The bottom panel shows the total area of NOAA regions scaled to the total area of SMART regions. In fact, the NOAA area is lower by a factor of $\sim$$50$, since only the low-intensity area of sunspots is summed, while the area of extended magnetic features is recorded in SMART detections. Number and area are the only two feature properties which can be directly compared, as NOAA data do not contain any magnetic property measurements.

\begin{figure}[!ht]
\begin{center}
\includegraphics*[width=12cm,angle=0]{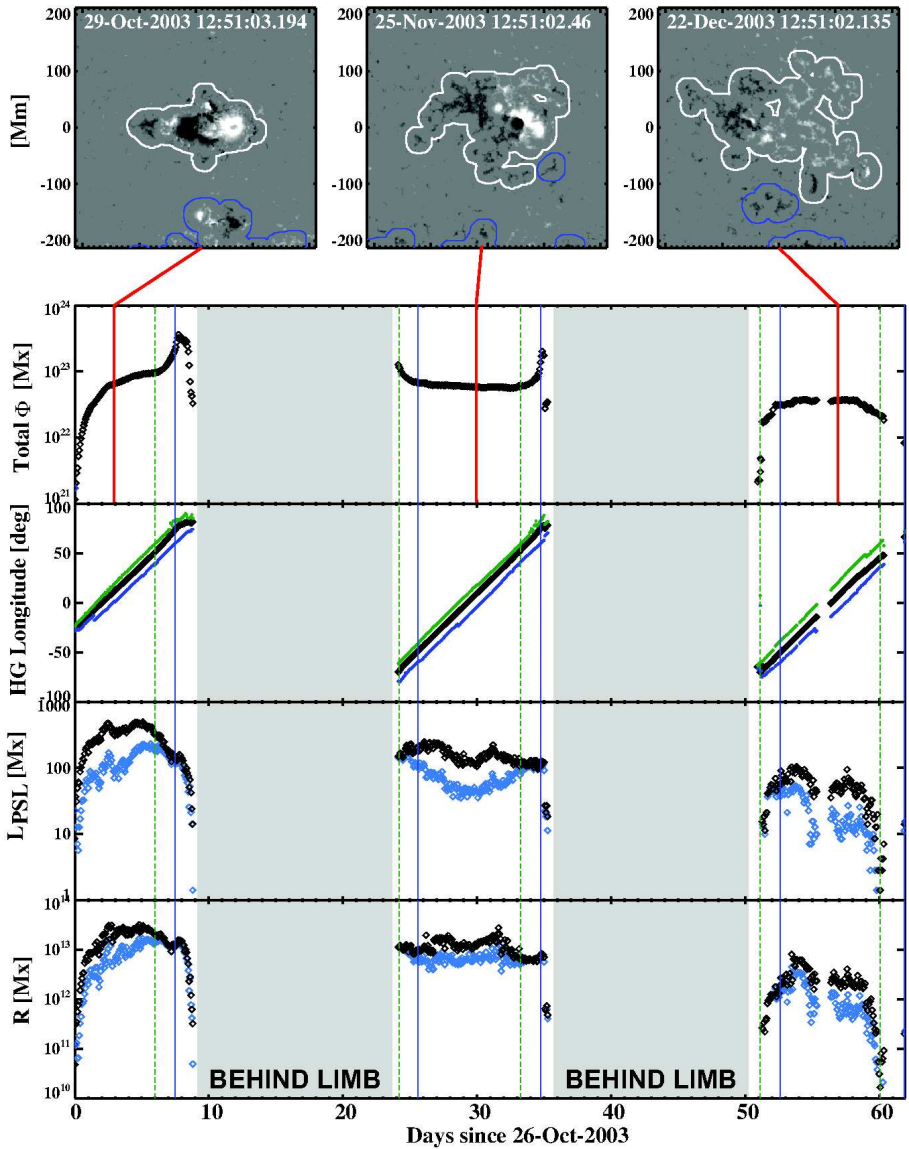}
\end{center}
\caption{Tracking of 20031026.MG.11 as it rotates around the Sun from 26 October to 26 December 2003.}\label{ar_track}
\end{figure}

The determination of the magnetic properties of a feature is affected by MDI magnetogram noise levels, calibration, strong field saturation, and LOS effects. The feature detection itself is generally not affected by these phenomena, however. The instrument noise threshold of MDI is nominally $\pm20$~G \citep{Sher95}. This is smoothed by the gaussian convolution, and the segmentation threshold of $\pm70$~G is well above this. For magnetic property calculations, a gaussian convolution is not used, so noise contributes $20$~G to the uncertainty of pixel values above the QS threshold of $70$~G. For SMART region 20031026.MG.11 observed at disk center on 25 November 2003, which is found to have a $A_{tot,t,i}$ of $3.8\times10^{4}$~Mm$^2$ and a $\Phi_{uns,t,i}$ of $5.9\times10^{22}$~Mx, the uncertainty is $7.9\times10^{21}$~Mx, or $5\%$.

Some calibration issues with the MDI data used by SMART are discussed in \citet{WZZ09}. It was found that the 2008 calibration of level 1.8 data has been partially corrected, in that it does not suffer from a disk center-to-limb variation like the 2007 calibration. However, MDI may largely underestimate the magnetic field as the ratio of MDI values to those retrieved from \emph{Hinode}/Solar Optical Telescope data was found to be $\sim$$0.7$. This does not affect feature detections since the effect is consistent throughout the data set, but could contribute a considerable error of $\sim$$30\%$ for any magnetic field or flux measurements. 

Strong magnetic field saturation in MDI data is discussed in \citet{LiuNS07}. It is estimated that this phenomenon occurs in $\sim$$5\%$ of ARs, in which the magnetic field measurements in the umbral areas of very strong sunspots behave non-linearly. In extreme cases, the umbra may appear to have a smaller magnetic field than the surrounding penumbra. In reality, the field should continue to increase in the umbra, but in level 1.8 data showing NOAA 9002 at disk center, saturation is clearly observed at $\sim$$3000$~G. Feature boundaries are not affected because saturation only occurs for very strong sunspot umbrae, although the derived magnetic properties of features which include strong sunspots will be underestimated. 

\begin{figure}[!ht]
\begin{center}
\includegraphics*[width=12cm,angle=0]{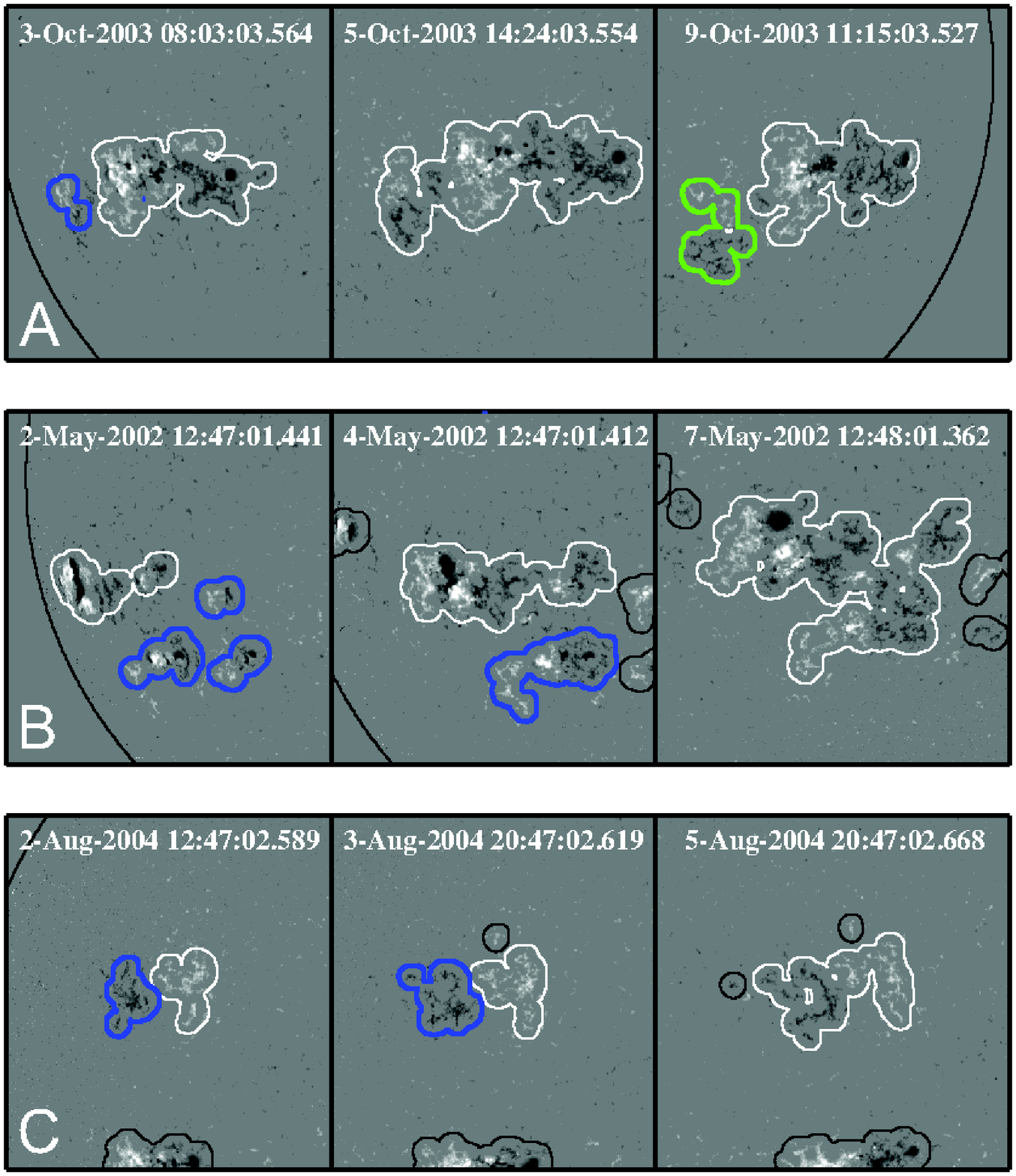}
\end{center}
\caption{Feature detection and tracking cases which diverge from NOAA. A) Two bipolar regions join and subsequently fragment. B) Several small bipolar regions merge into an AR complex. C) A bipolar region is first detected as two unipolar features and then as a single bipolar region.}\label{detectcases}
\end{figure}

LOS effects occur when features are not observed at disk center. To estimate the effects of this we model a circular spot with an area of $1.6\times10^4$~Mm$^2$ progressing to the edge of the solar disk. 
The LOS area is measured at longitude increments of $3$~degrees and modified by the SMART cosine area correction. This is compared to the disk center area of the spot, resulting in an over correction of $\sim$$3\%$ when the centroid reaches $60$~degrees longitude. This error depends on morphology and will be more acute for complex feature boundaries. The over correction increases quickly to $\sim$$40\%$ as the feature is tracked toward the limb.

An example of the SMART method of feature tracking and cataloging is shown in Figure~\ref{ar_track}. Region 20031026.MG.11 is tracked from 26 October 2003 to 26 December 2003. The AR rotates beyond the west limb and is detected again upon returning at the east limb twice. Although the AR is tracked to subsequent solar rotations its catalog name remains the same when it returns. NOAA first detects this AR on 28 October 2003 designating it as NOAA 10488. When the region returns it is designated a new region number, NOAA 10507 and is renamed upon the second return as NOAA 10525. SMART's persistent naming through multiple rotations allows independent measurements of the same feature to be grouped into a single time plot.

The top panels in Figure~\ref{ar_track} show MDI magnetograms of the region (clipped at $\pm1000$~G) on three different dates, with the extracted AR outlined by a thick white contour (other detections are outlined in blue). A connecting red line shows where each falls on the timeline below. The remaining panels show, from top to bottom, time series of total unsigned flux (${\Phi}_{uns,t,i}$), heliographic longitude ($HG_{pos,t,i}$), PSL length ($L_{PSL,t,i}$), and R value ($R_{t,i}$) extracted from 20031026.MG.11. Vertical dotted green (blue) lines denote crossings at $\pm60$~degrees of the leading (trailing) edge of the feature; in the second time plot, the green (blue) curve tracks this leading (trailing) edge in time. In the plot of PSL length, the black curve sums the length of all detected PSL segments ($L_{PSL,t,i}$), while the light-blue curve sums those having a gradient above $50$~G~Mm$^{-1}$ ($L_{sg,t,i}$). Finally, the plot of R-value shows $R^{*}_{t,i}$ in black and $R_{t,i}$ in blue.

The stability of the algorithm is estimated using the plot of ${\Phi}_{uns,t,i}$ between days 25.6 (20 November 14:24 UT) and 33.7 (28 November 16:48 UT). A quadratic fit is subtracted to remove the long timescale variation, resulting in an array of residuals. The two-sigma error of the residuals is determined to be $2.1\times10^{21}$~Mx or $3\%$ around the mean of ${\Phi}_{uns,t,i}$. The stability estimate is particular to this example, as cases such as those shown in Figure~\ref{detectcases} could result in much larger short timescale variation.

There are several recurrent SMART feature tracking cases which diverge from what would be expected of NOAA (Figure~\ref{detectcases}). The SMART tracking algorithm allows features to converge and split apart. However, there may be side-effects, such as when a fragment separates from a larger feature and is given a new catalog name, due to the centroids of the two being greater than the tracking association threshold (top row). Also, an active region complex  may be detected when there are multiple strong field ARs in close proximity (middle row). Finally, a bipolar region which is significantly disjointed and weak may not be properly grouped into a single region (bottom row). Here we see an example where each polarity is detected as a separate region. As this work is designed to aid in flare forecasting, many examples of each of these cases may be studied to determine if they possess unexpected flaring properties. Also, their evolution maybe studied by tracking the features from first emergence. The frequency of occurrence for these special cases can be estimated using the data and analysis of Figure~\ref{noaa_comp}: when $N_{NOAA}$ is greater than $N_{SMART}$ SMART is likely grouping regions into AR complexes (or identifying NOAA ARs as EF or DF), and when $N_{SMART}$ is greater than $N_{NOAA}$ SMART may be detecting individual unipolar features when NOAA groups them into bipolar regions.


\section{Conclusions}\label{disc_conc}

\indent The SMART algorithm allows one to monitor ARs on the solar disk in near-realtime and perform extensive studies on AR magnetic properties. SMART is unique among automated AR extraction algorithms in that it allows the temporal analysis of magnetic properties from birth and through multiple solar rotations. Future work will include the analysis of trends in AR evolution over the solar cycle. This is a largely untouched subject that begs important questions, such as whether ARs are born destined to flare or randomly evolve to become flare-active. This may also provide new insights into the behavior of the solar dynamo.

\indent Previous algorithms include some of the functions performed by the
SMART algorithm, such as feature and magnetic parameter extraction. However,  new utilities are incorporated into the SMART code, such as day-to-day and multiple rotation feature tracking. Extensive AR properties such as area ($A_{tot,t,i}$) and total magnetic flux ($\Phi_{uns,t,i}$) are determined, as are intensive properties such as the maximum magnetic field ($B_{max,t,i}$) and statistical moments ($\mu,\sigma^2,\gamma,\kappa$). Some algorithms, including \citet{LaBon07} only detect the largest regions, while others like \citet{CoQa09} only detect ARs with sunspots in white-light images. All current algorithms track ARs using visually identified NOAA specifications. The SMART algorithm is independent from these specifications and needs no human intervention to detect and track ARs. Additionally, it utilizes an improved feature cataloging system which incorporates the date of first detection and the feature type.


\indent The SMART algorithm will be used to create a comprehensive catalog of features present in magnetograms covering the entirety of solar cycle 23 and will be adapted to use \emph{SDO}/Helioseismic and Magnetic Imager data. A pipeline version of the algorithm will output detections for inclusion in the Heliophysics Event Knowledgebase\footnote{See: http://www.lmsal.com/helio-informatics/hpkb/index.html}. Additionally, it will form part of HELIO. In this application, ARs tracked using SMART will be associated with a chain of features and events propagating throughout the heliosphere, such as EUV loops, flares, CMEs, magnetic disturbances and storms detectable in Earth's aurorae and ground-based magnetometer data, as well as distant particle instruments such those on the Voyager and Mercury Surface, Space Environment, Geochemistry, and Ranging (\emph{MESSENGER}) spacecraft. 

\indent The magnetic properties of ARs retrieved by the SMART algorithm will also be used for flare forecasting. While the magnetic complexity of ARs is known to be an important predictor of flare activity \citep{SZT00,Shj07,McAt05b,Con08}, recent work by \citet{Wel09} shows  that extensive magnetic properties outperform intensive properties as predictors of AR flare activity. One of SolarMonitor's current flare-forecasting algorithms assumes Poisson statistics \citep{Mn01,Wh01,Gal02} and relies on historical  flaring rates from 1988 to 1996 for each McIntosh sunspot classification \citep{McIn90}. This will be superseded by a statistical forecasting algorithm that makes use of extensive AR magnetic properties determined by SMART.

Any forecasting algorithm which makes use of magnetic properties output by SMART will need to take into account several sources of error. Random errors including magnetogram noise and algorithm stability for the example presented in Section~\ref{results} result in an error of $\pm5\%$ and $\pm3\%$ in $\Phi_{uns,t,i}$, respectively. This will not affect the forecasting potential of properties involving ${\Phi}_{uns,t,i}$ for a sufficiently large sample of regions. Calibration errors in MDI result in an underestimate of the true magnetic field on average by $\sim$$30\%$. If the forecasting training set and test samples both exhibit this error, the prediction result will not be affected. However, for physical studies of energetics this must be taken into account. Finally, LOS effects which occur as regions approach the limb cause large measurement errors past $60$~heliographic degrees from disk center, which limits the potential forecasting range of this algorithm. 

\section{Acknowledgements}

This research is supported by ESA/PRODEX and a grant from the EC Framework Programme 7 (HELIO). RTJMcA (FP6) and DSB (FP7) are Marie Curie Fellows at TCD. We would like to thank the SOHO team for making both their data and analysis software publicly available and to acknowledge the participants of the first `Forecasting the All-Clear' meeting (April 22-24, 2009) who provided helpful comments and insights upon the presentation of this work. We would also like to show our appreciation to the two anonymous referees whose comments helped to improve this paper.


\clearpage

\end{document}